\documentstyle[12pt]{article}

\def\tomega{\tilde{\omega}}
\def\ket#1{| #1 \rangle}
\def\bra#1{\langle #1 |}

\begin{document}
\centerline{\large\bf Photon Frequency Entanglement Swapping} 

\vskip 3mm
\centerline{S.N.Molotkov and S.S.Nazin}
\vskip 2mm
\centerline{\sl\small Institute of Solid State Physics of Russian Academy of
Sciences}
\centerline{\sl\small Chernogolovka, Moscow District, 142432 Russia}
\centerline{E-mail: molotkov@issp.ac.ru, nazin@issp.ac.ru}
\vskip 3mm
\begin{abstract}
We propose a simple non-linear crystal based optical scheme for
experimental realization of the frequency entanglement swapping
between the photons belonging to two independent biphotons.
\end{abstract}

Entangled states allowed by quantum theory result in purely quantum 
non-local correlations which do not occur in classical physics.
The correlations of that kind underlie the well-known EPR effect 
\cite{EPR} which was most clearly formulated by Bohm \cite{Bohm}  
using the entangled singlet state of a pair of spin-1/2 particles. 
Five years ago Bennett {\sl et al} \cite{teleport} proposed an algorithm
for the teleportation of an arbitrary unknown state of a two-level
quantum system also based on employing the entangled state of an EPR-pair.
After the algorithm is performed, the initial state to be teleported
is ``swapped'' (to within a unitary transformation which is completely
specified by the data transmitted via classical communication channel 
to the user receiving the teleported state) into the state of a particle
belonging to the auxiliary EPR pair; the swapping is achieved by performing
a joint measurement involving both the initial particle and the
other particle in the EPR pair (the very idea of ``entanglement swapping'' 
was first introduced in \cite{Zukowski1993} when discussing the possibility
of experimental realization of producing an entangled state of
a pair of particles which have never interacted in the past
and was further developed in \cite{Zukowski1995}; its 
multi-particle extension was given in Ref.\cite{Bose}). 
Experimentally the quantum teleportation
was realized in Refs.\cite{ExpTel_1,ExpTel_2}. Teleportation of the state 
of quantum system described by continuous dynamical variables (e.g.,
position and momentum of a one-dimensional particle) is discussed
in Refs.\cite{Vaidman,BraunKimble,Molotkov_1,Molotkov_2}.
Due to non-local correlations, the entangled states can also
be used in quantum cryptography in the key distribution schemes
\cite{Phoenix,repeaters}. 

Recent paper \cite{EntSwap} experimentally demonstrated the possibility
of preparation of an entangled state of two photons which never 
interacted in the past. In that scheme, the entanglement involved
two possible (``horizontal'' and ``vertical'') polarization states
of the photons. In the present paper we propose a conceptually very 
simple scheme for experimental realization
where the entanglement swapping occurs between the frequency states
of the photons initially belonging to two independent biphotons
(it is interesting to note that the frequency entangled states
of two photons demonstrate the EPR effect with respect to the
time and energy observables as discussed in Ref.\cite{Klyshko}).
Although all proposed realizations of quantum teleportation
of a discrete variable (e.g., photon polarizations) and continuous
variable employ the non-linear crystals to produce entangled photon pairs,
it should be emphasized that the type of experiment we wish to propose 
here does not involve any path interference schemes as discussed in
Ref.\cite{Zukowski1993} and is in some sense analogous to the method
of teleportation of a continuous variable proposed 
in Refs.\cite{Molotkov_1,Molotkov_2}.

Suppose we are given two biphotons $\ket{\Phi}$ and $\ket{\Psi}$:
\begin{equation}
\ket{\Phi}=\int d\omega_1 f(\omega_1)\ket{\omega_1} 
\otimes\ket{\Omega_I - \omega_1},\qquad 
\ket{\Psi}=\int d\omega_3 g(\omega_3)\ket{\omega_3} 
\otimes\ket{\Omega_{II} - \omega_3}; 
\label{biphot}
\end{equation}
here $\Omega_I$ and $\Omega_{II}$ are some fixed frequencies,
$\ket{\omega}$ is the single-photon monochromatic Fock state
of the radiation field and we omit the photon polarization and spatial
(wavevector direction) degrees of freedom, assuming that each photon 
follows his own arm of the optical scheme.
The integration in (\ref{biphot}) is performed over all frequencies
corresponding to positive frequencies of the ket-states.
For reasons of brevity, we shall further refer to the photons
$\ket{\omega_1}$ and $\ket{\Omega_I - \omega_1}$ in the biphoton 
$\ket{\Phi}$ as the first and second photon, respectively,
while to the photons $\ket{\omega_3}$ and $\ket{\Omega_{II} - \omega_3}$ in the
biphoton $\ket{\Psi}$ as the third and fourth photon. 
The frequency entanglement of the states (\ref{biphot}) can be 
verified by joint measurements of the frequencies (energies) 
of the photons belonging, for example, to $\ket{\Phi}$. 
Thus, placing in the paths of these photons narrow-band photodetectors
tuned to frequencies $\omega^{(1)}$ and $\omega^{(2)}$ for the first 
and second photons, respectively, one would only observe simultaneous 
firing of these detectors if the following condition is satisfied:
\begin{equation}
 \omega^{(1)} + \omega^{(2)} = \Omega_I.
\end{equation}

Note that if the frequency dependence of $f(\omega)$ in
(\ref{biphot}) is neglected the photons belonging to the biphoton
$\ket{\Phi}$ exhibit a very simple temporal correlation. Namely, 
if both photons are registered by wide-band photodetectors then
the photodetectors will fire at exactly the same time (if they are
equidistant from the biphoton source). Indeed, similarly to
Ref.\cite{Holevo}, the identity resolution corresponding to the joint
measurement of the registration times of the first and second photons
$t_1$ and $t_2$ can be written as 
\begin{equation}
  M(dt_1,dt_2) = \frac{dt_1 dt_2}{(2\pi)^2}
 \int\!\! \int\!\! \int\!\! \int 
     d\tomega_1 d\tomega_1' d\tomega_2 d\tomega_2'
       e^{i(\tomega_1 - \tomega_1')t_1}
       e^{i(\tomega_2 - \tomega_2')t_2}
           \ket{\tomega_1}\otimes\ket{\tomega_2}
           \bra{\tomega_1'}\otimes\bra{\tomega_2'}.
\end{equation}
Hence the joint probability density distribution for observing the values
$t_1$ and $t_2$ within the intervals ($t_1, t_1 + dt_1$) and 
($t_2, t_2+dt_2$) is
\begin{equation}
 {\rm Pr}(dt_1,dt_2) = {\rm Tr}\{ \ket{\Phi}\bra{\Phi}\, M(dt_1,dt_2)\}.
\end{equation}
Straightforward calculations reveal that if $f(\omega_1)$ in (\ref{biphot}) 
does not depend on frequency then
\begin{equation}
 {\rm Pr}(dt_1,dt_2) = \frac{dt_1 dt_2}{(2\pi)^2}
 \int\!\! \int\  
    d\tomega_2 d\tomega_2'
       e^{i(\tomega_2' - \tomega_2)(t_1-t_2)}
        f(\Omega - \tomega_2')
        f^*(\Omega - \tomega_2) 
\sim
     \delta(t_1 - t_2),
\end{equation}
since in that case the integration over
$d\tomega_2 d\tomega_2'$ can be replaced by integration over
$d(\tomega_2 - \tomega_2')$ and $d(\tomega_2 + \tomega_2')$,
and the integration over $d(\tomega_2 - \tomega_2')$ yields $\delta(t_1 - t_2)$.

It is obvious that to create entanglement between the first and fourth photons,
i.e. to perform entanglement swapping between the photons belonging to the
biphotons $\ket{\Phi}$ and $\ket{\Psi}$ one should modify the 
initial state
\begin{equation}
\ket{\Phi} \otimes \ket{\Psi} =   
\int d\omega_1 d\omega_3 f(\omega_1)g(\omega_3)
\ket{\omega_1} \otimes\ket{\Omega_I - \omega_1} \otimes 
\ket{\omega_3} \otimes\ket{\Omega_{II} - \omega_3} 
\label{initial}
\end{equation}
in such a way that only the tensor products involving the states
$\ket{\omega_1}$ and $\ket{\Omega_{II} - \omega_3}$ 
with somehow correlated frequencies which cannot independently
take arbitrary values are left under the integral sign.
To achieve this goal one can, for example, project the initial state on the 
subspace spanned by the states characterized by an arbitrarily chosen
constant sum $\Omega_{III}$ of the second and third photon frequencies. 
Starting from the standard resolution of identity
\begin{equation}
 I=\int d\Omega \int d\omega \ket{\omega} \otimes\ket{\Omega - \omega}
                \bra{\omega} \otimes \bra{\Omega - \omega}, 
\label{FullResolution}
\end{equation}
the indicated projection can be realized in the limit 
$\Delta\Omega \to 0$ by performing a measurement
which is formally described by the simplest identity resolution
\begin{equation}
 P_{\Delta\Omega}+P'_{\Delta\Omega}=I,
\end{equation}
where the orthogonal projector $P_{\Delta\Omega}$ is defined as
\begin{equation}
 P_{\Delta\Omega}=\int_{\Omega_{III}}^{\Omega_{III}+\Delta\Omega} d\Omega
                  \int d\omega 
      \ket{\omega} \otimes\ket{\Omega - \omega}
      \bra{\omega} \otimes \bra{\Omega - \omega}. 
\label{proj}
\end{equation}

Indeed, if the measurement of the projector $P_{\Delta\Omega}$ 
(whose eigenvalues, as for any projector, are 0 and 1) yields
the value of 1, it is easily verified that the state (\ref{initial})
transforms to the state
\begin{displaymath}
\ket{\Xi} =   
\int_{\Omega_{III}}^{\Omega_{III}+\Delta\Omega} d\Omega
\int d\omega_1 f(\omega_1)g(\Omega - \Omega_{I} + \omega_1)\times
\end{displaymath}
\begin{equation}
\ket{\omega_1} \otimes\ket{\Omega_I - \omega_1} \otimes 
\ket{\Omega - (\Omega_I - \omega_1)} \otimes
\ket{\Omega_{II} + \Omega_{I} - \Omega - \omega_1}, 
\label{XiProj}
\end{equation}
showing that in the limit $\Delta\Omega \to 0$ 
the first and fourth photons become frequency-entangled.
On the other hand, if the measurement outcome is 0, the initial state
will be modified in such a way that there will be no simple
correlation between the frequencies of the first and fourth photon. 

Consider now the possibility of experimental realization of the
frequency entanglement swapping between the two biphotons (Fig.1).
Suppose we have two non-linear crystals (I and II) with non-zero 
second order susceptibility $\chi$ so that the radiation field 
Hamiltonian in the crystals contains the terms \cite{Mandel,Milloni}
\begin{equation}    
H_1(t)=\chi\int d{\bf x}E^{(+)}({\bf x},t)E^{(-)}({\bf x},t)
E^{(-)}({\bf x},t) + \rm{h.c.},
\end{equation}
where $E$ is the electric field operator and all unimportant
constants are included into the definition of the
susceptibility $\chi$ which is usually assumed to be frequency independent;
the superscripts $+$ and $-$ at the electric field operator label 
the positive and negative frequency parts, respectively \cite{Campos}: 
\begin{equation}    
E^{(\pm)}({\bf x},t)= \frac{1}{\sqrt{2\pi}}
\int_{0}^{\infty}d\omega e^{i(\mp\omega t-{\bf kx})}
\hat{a}^{\mp}(\omega), 
\label{PosNegFreq}
\end{equation}
where $a^+$ and $a^-$ are the creation and anihilation oprators,
so that $a^+(\omega)\ket{0} = \ket{\omega}$,
$\ket{0}$ being the vacuum state of the radiation field.
Bearing in mind that we shall be interested in the processes of the
decay of incident photons into a pair of photons (in crystals I and II)
and merging of two incident photons into a single one (in crystal III), 
we have
\begin{equation}   
H_1(t)=\frac{\chi}{(2\pi)^{3/2}}
\int\!\! \int\!\! \int
d\omega_1d\omega_2 d\omega_{in}
e^{ it( \omega_1+\omega_2-\omega_{in} ) }
\ket{\omega_1}\otimes \ket{\omega_2} \bra{\omega_{in}}
\int_{vol}d{\bf x} e^{-i{\bf x}({\bf k_1+k_2-k_{in}})}+\rm{h.c.}
\label{H1}
\end{equation}
where the subscript {\sl in} refers to the incident photon and
while subscripts 1 and 2 label the two generated photons
(see Fig.1); in the hermitian conjugated part of (\ref{H1})
the subscript {\sl in} should be replaced by {\sl out}.
In the second integral the integration is performed over the entire
crystal yielding the $\delta$-symbol with respect to the photon momenta,
i.e. the condition ${\bf k_1+k_2}={\bf k_{in}}$ is satisfied. 

After the passage through the nonlinear crystal the radiation field state is 
\begin{equation}  
\ket{\Phi}_{out}=S(t)\ket{\Phi}_{in},
\end{equation}
where the $S$-matrix can be written as an expansion in powers of $\chi$:
\begin{equation}    
S(t)=e^{i\int_{-\infty}^{t}H_1(t')dt'}=1+S^{(1)}+S^{(2)}+\ldots
\end{equation}
To the first order in $\chi$ 
\begin{equation}  
S^{(1)}=i\chi
\int d\omega_1  
\int d\omega_{in}
\ket{\omega_1}_1\otimes 
\ket{\omega_{in}-\omega_1}_2 \,\,\,_{in}\bra{\omega_{in}} + \rm{h.c.}, 
\end{equation}
because the upper limit of the integration over time in the exponent in $S$ can be set to
$\infty$ since actually the photon wave packet travels through the crystal
in a finite time. To within an unimportant normalization constant,
the component in the output state which we are interested in
(the one arising from $S^{1}$) is
\begin{equation}   
\ket{\Phi_{EPR}}=\chi
\int_{0}^{\infty}d\omega\ket{\omega}_1\otimes\ket{\Omega-\omega}_2,
\end{equation}
which is exactly the required biphoton. Thus, irradiation
of the crystals I and II by monochromatic photons with frequencies
$\Omega_I$ and $\Omega_{II}$ results in the creation of biphotons
(\ref{biphot}), where $f\sim \chi_I, \quad g\sim \chi_{II}$, i.e. 
the relative frequency of positive outcomes (generation of a biphoton) 
being proportional to the square of the small non-linear
susceptibility $\chi$ of the crystals. 
Let further the second and third photons with the frequencies
$\omega_2$ and $\omega_3$, respectively, simultaneously hit 
the non-linear crystal III where they can merge with certain probability
into a single photon with frequency $\omega' = \omega_2 + \omega_3$.
Finally, the output of crystal III is registered
by a narrow-band photodetector tuned to the frequency $\Omega_{III}$. 
Such a photodetector together with a nonlinear crystal placed in front of it
serve as a device realizing the measurement of a physical quantity 
defined by the operator $P_{\Delta\Omega}$ from (\ref{proj}) 
in the limit $\Delta\Omega \to 0$, the only difference being
that the firing of the photodetector implies vanishing of the
registered photon.

The photodetector $P$ fires with the probability which is proportional,
among other factors, to $\chi_{III}^2$ and in that case the photons 2 and 
3 vanish while the rest two photons (first and fourth) are left in the
state $|\Xi'\rangle$ which can be obtained from (\ref{XiProj}) by simply
cancelling the second and third photon:
\begin{equation}
|\Xi'\rangle =   
\int d\omega_1 f(\omega_1)g(\Omega_{III} - \Omega_{I} + \omega_1)
\ket{\omega_1} \otimes
\ket{\Omega_{II} + \Omega_{I} - \Omega_{III} - \omega_1}. 
\label{XiCancelProj}
\end{equation}
Thus, the first and fourth photons become frequency-entangled.
To experimentally verify their entanglement, one can register them
with two narrow-band photodetectors $P_1$ and $P_4$ (see Fig. 1)
tuned to the frequencies
$\Omega_1$ and $\Omega_4$. In that case a simultaneous firing
of these photodetectors (of course, we consider only the cases when
the photodetector $P$ has fired since otherwise we cannot be sure that
the second and third photons merged in a single photon with the
required frequency $\Omega_{III}$) should only be observed
if the frequencies $\Omega_1$ and $\Omega_4$ satisfy the condition
\begin{equation}
 \Omega_1 + \Omega_4 = \Omega_I + \Omega_{II} - \Omega_{III}.
\end{equation}

The work was supported by the Russian Fund for Basic Research
(grants No 96-02-19396 and 98-02-16640)
and the National program ``Advanced technologies in micro- and 
nanoelectronics.

\newpage 

\begin{figure}[tb]
\caption{Schematics of the proposed experiment on the entanglement
swapping with the two biphotons produced in non-linear crystals
by monochromatic photons with the frequencies $\Omega_I$ and 
$\Omega_{II}$.}
\end{figure}

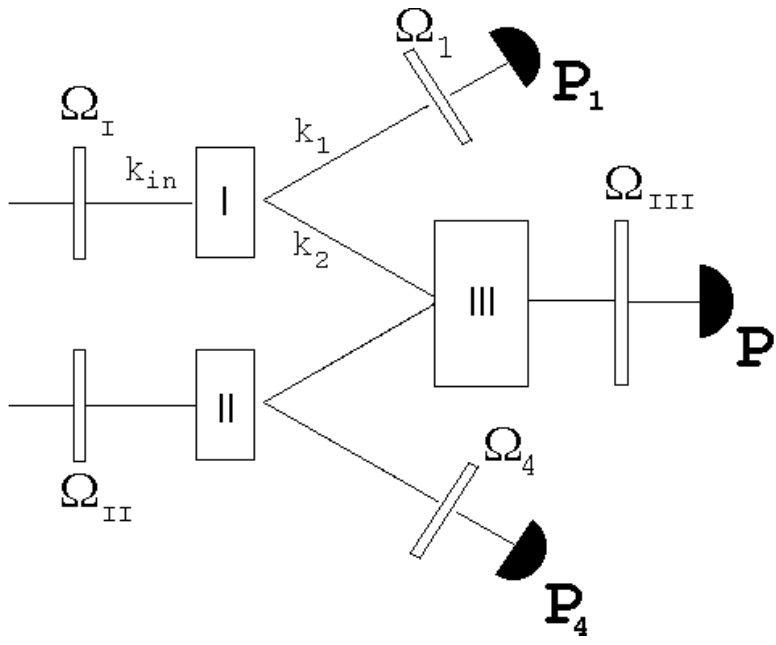
\end{document}